# ASYMPTOTIC ANALYSIS OF DOWNLINK MISO SYSTEMS OVER RICIAN FADING CHANNELS

*Hugo Falconet*[⋆], *Luca Sanguinetti*[†‡], *Abla Kammoun*[+], *Merouane Debbah*[⋆†]

[⋆]Mathematical and Algorithmic Sciences Lab, Huawei France, Paris, France
[†]Large Networks and System Group (LANEAS), CentraleSupélec, Université Paris-Saclay, Gif-sur-Yvette, France
[‡]Dipartimento di Ingegneria dell'Informazione, University of Pisa, Pisa, Italy
[+]Electrical Engineering Department, King Abdullah University of Science and Technology, Thuwal, Saudi Arabia

## ABSTRACT

In this work, we focus on the ergodic sum rate in the downlink of a single-cell large-scale multi-user MIMO system in which the base station employs $N$ antennas to communicate with $K$ single-antenna user equipments. A regularized zero-forcing (RZF) scheme is used for precoding under the assumption that each link forms a spatially correlated MIMO Rician fading channel. The analysis is conducted assuming $N$ and $K$ grow large with a non trivial ratio and perfect channel state information is available at the base station. Recent results from random matrix theory and large system analysis are used to compute an asymptotic expression of the signal-to-interference-plus-noise ratio as a function of the system parameters, the spatial correlation matrix and the Rician factor. Numerical results are used to evaluate the performance gap in the finite system regime under different operating conditions.

## 1. INTRODUCTION

Large-scale multiple-input multiple-output (MIMO) systems (also known as massive MIMO systems) are considered as one of the most promising technology for next generation wireless communication systems [1–3] because of their considerable spatial multiplexing gains. The use of large-scale MIMO systems is beneficial not only in terms of coverage and spectral efficiency but also in terms of energy-saving [4–6]. In this complex system model, a number of practical factors such as correlation effects and line-of-sight (LOS) components need to be included, which occur due to the space limitation of user equipments (UEs) and the densification of the antenna arrays resulting in a visible propagation path from the UEs, respectively. For typical systems of hundreds of antennas and tens of UEs, even computer simulations become challenging, which makes performance analysis of large-scale MIMO systems an important and a new subject of research.

In this work, we consider the downlink of a single-cell large-scale MIMO system in which the base station (BS), equipped with $N$ antennas, makes use of regularized zero-forcing (RZF) precoding to communicate with $K$ single-antenna UEs. In particular, we are interested in evaluating the ergodic sum rate of the system when a power constraint is imposed at the BS. The analysis is conducted assuming that $N$ and $K$ grow large with a non trivial ratio under the assumption that perfect channel state information is available at the BS. Differently from most of the existing literature [7–11], we consider a spatially correlated MIMO Rician fading model, which is more general and accurate to capture the fading variations when there is a LOS component. Compared to the Rayleigh fading channel, a Rician model makes the asymptotic analysis of large-scale MIMO systems much more involved. To overcome this issue, recent results from random matrix theory and large system analysis [11–13] are used to compute an asymptotic expression of the signal-to-interference-plus-noise ratio (SINR), which is eventually used to approximate the ergodic sum rate of the system. As shall be seen, the results are found to depend only on the system parameters, the spatial correlation matrix and the Rician factor. As a notable outcome of this work, the above analysis provides an analytical framework that can be used to evaluate the performance of the network under different settings without resorting to heavy Monte Carlo simulations and to eventually get insights on how the different parameters affect the performance.

The main literature related to this work is represented by [7, 10, 14–16]. Tools from random matrix theory are used in [7] to compute the ergodic sum rate in a single-cell setting with Rayleigh fading and different precoding schemes while the multicell case is analyzed in [10]. In [14], the authors investigate a LOS-based conjugate beamforming transmission scheme and derive some expressions of the statistical SINR under the assumption that $N$ grows large and $K$ maintain fixed. In [15], the authors study the fluctuations of the mutual information of a cooperative small cell network operating over a Rician fading channel under the form of a central limit theorem and provide explicit expression of the asymptotic variance. In [16], a deterministic equivalent of the ergodic sum rate and an algorithm for evaluating the capacity achieving input covariance matrices for the uplink of a large-scale MIMO are proposed for spatially correlated MIMO channel with LOS components. The analysis of the uplink rate with both zero-forcing and maximum ratio combining receivers is performed in [17].

The following notation is used throughout this work. The superscript $^H$ stands for the conjugate transpose operation. The operator $\mathrm{Tr}\mathbf{X}$ denotes trace of matrix $\mathbf{X}$ whereas $\mathbf{X}_{[k]}$ indicates that the $k$th column is removed from matrix $\mathbf{X}$. The $N \times N$ identity matrix is denoted by $\mathbf{I}_N$ whereas $\mathbf{X} = \mathrm{diag}\{x_1, \ldots, x_N\}$ is used to denote a $N \times N$ diagonal matrix of entries $\{x_n\}$. A random variable $x$ is a standard complex Gaussian variable if $\sim \mathcal{CN}(0, 1)$.

## 2. SYSTEM AND CHANNEL MODEL

We consider the downlink of a network in which $K$ UEs are served by a single BS equipped with $N$ antennas. The signal $y_k$ received

This research has been supported by the ERC Starting Grant 305123 MORE and by the HUAWEI project RMTin5G.

by UE $k$ takes the form [7]

$$y_k = \mathbf{h}_k^H \sum_{k=1}^{K} \mathbf{g}_k s_k + n_k \quad \forall k \in \{1, \ldots, K\} \quad (1)$$

where $\mathbf{h}_k \in \mathbb{C}^N$ is the random channel from the BS to user $k$, $\mathbf{g}_k \in \mathbb{C}^N$ is the precoding vector associated to UE $k$, and $n_k \sim \mathcal{CN}(0, \sigma^2)$ accounts for thermal noise. For analytic tractability, we assume that the BS is able to acquire perfect channel state information from the uplink pilots. The RZF scheme is used as precoding technology [7]. Therefore, the precoding matrix $\mathbf{G} = [\mathbf{g}_1 \ldots \mathbf{g}_K] \in \mathbb{C}^{N \times K}$ takes the form

$$\mathbf{G} = \xi \left( \mathbf{H}\mathbf{H}^H + \lambda K \mathbf{I}_N \right)^{-1} \mathbf{H} \quad (2)$$

where $\mathbf{H} = [\mathbf{h}_1, \ldots, \mathbf{h}_K] \in \mathbb{C}^{N \times K}$ is the aggregate channel matrix, $\lambda > 0$ is the so-called regularization parameter and $\xi$ is chosen so as to satisfy the average total power constraint given by $\mathrm{Tr}\mathbf{G}^H \mathbf{G} = P_T$ where $P_T > 0$ denotes the total available transmit power. Therefore, it follows that

$$\xi^2 = \frac{P_T}{\mathrm{Tr}\mathbf{H}^H \left( \mathbf{H}\mathbf{H}^H + \lambda K \mathbf{I}_N \right)^{-2} \mathbf{H}}. \quad (3)$$

Under the assumption of Gaussian signaling, i.e., $s_k \sim \mathcal{CN}(0, 1)$ and single user detection with perfect channel state information at the receiver, the SINR $\gamma_k$ of user $k$ takes the form

$$\gamma_k = \frac{|\mathbf{h}_k^H \mathbf{g}_k|^2}{\mathbf{h}_k^H \mathbf{G}\mathbf{G}^H \mathbf{h}_k - |\mathbf{h}_k^H \mathbf{g}_k|^2 + \sigma^2}. \quad (4)$$

The rate $r_k$ of UE $k$ is given by $r_k = \log_2(1 + \gamma_k)$ whereas the ergodic sum rate is defined as

$$r_E = \sum_{k=1}^{K} \mathbb{E}\left[ \log_2(1 + \gamma_k) \right] \quad (5)$$

where the expectation is taken over the random channels $\{\mathbf{h}_k\}$. We assume that

$$\mathbf{h}_k = \sqrt{\beta_k} \mathbf{w}_k \quad (6)$$

where $\beta_k$ accounts for the large scale channel fading of UE $k$ and $\mathbf{w}_k \in \mathbb{C}^N$ is the small scale fading channel. The latter is modelled as

$$\mathbf{w}_k = \sqrt{\frac{1}{1+\rho}} \mathbf{\Theta}^{1/2} \mathbf{z}_k + \sqrt{\frac{\rho}{1+\rho}} \tilde{\mathbf{z}}_k \quad (7)$$

where $\mathbf{z}_k \in \mathbb{C}^N$ is assumed to be Gaussian with zero mean and unit covariance, i.e., $\mathbf{z}_k \sim \mathcal{CN}(\mathbf{0}_N, \mathbf{I}_N)$, and $\tilde{\mathbf{z}}_k \in \mathbb{C}^N$ is a deterministic vector. The scalar $\rho \geq 0$ is the Rician factor whereas the matrix $\mathbf{\Theta}^{1/2}$ is obtained from the Cholesky decomposition of $\mathbf{\Theta} \in \mathbb{C}^{N \times N}$, which accounts for the channel correlation matrix at the BS antennas. To make the problem analytically more tractable, we consider a system with common UE channel correlation matrix [7, 18, 19]. Although possible, the extension to the case in which UEs have different channel correlation matrices is mathematically much more involved and it is left for future research.

## 3. MAIN RESULT

We aim to exploit the statistical distribution of the channel $\mathbf{H}$ and the large dimensions of $N$ and $K$ to compute the deterministic approximation of $\gamma_k$, which will be eventually used to find an approximation of the ergodic sum rate. To begin with, we call

$$\tilde{\mathbf{D}} = \mathrm{diag}\left\{ \frac{\beta_1}{1+\rho}, \ldots, \frac{\beta_K}{1+\rho} \right\} \quad (8)$$

and compute the eigendecomposition of $\mathbf{\Theta}^{1/2}$ to obtain

$$\mathbf{\Theta}^{1/2} = \mathbf{U}^H \mathbf{D}^{1/2} \mathbf{U} \quad (9)$$

with $\mathbf{U} \in \mathbb{C}^{N \times N}$ being unitary. Then, we rewrite $\mathbf{H}$ as follows $\mathbf{H} = \sqrt{K} \mathbf{U}^H \mathbf{\Sigma}$ where $\mathbf{\Sigma}$ is defined as

$$\mathbf{\Sigma} = \frac{1}{\sqrt{K}} \mathbf{D}^{1/2} \mathbf{X} \tilde{\mathbf{D}}^{1/2} + \mathbf{A} \quad (10)$$

with $\mathbf{X} = \mathbf{U}\mathbf{Z}$, $\mathbf{A} = \mathbf{U}\tilde{\mathbf{Z}}$ and

$$\tilde{\mathbf{Z}} = \sqrt{\frac{1}{K}\frac{\rho}{1+\rho}} \left[ \sqrt{\beta_1} \tilde{\mathbf{z}}_1, \ldots, \sqrt{\beta_K} \tilde{\mathbf{z}}_K \right]. \quad (11)$$

Also, we assume the following grow rate of system dimensions:

**Assumption 1.** *The dimensions $N$ and $K$ grow to infinity at the same pace, that is:*

$$1 \leq \liminf N/K \leq \limsup N/K < \infty. \quad (12)$$

For technical reasons, the following reasonable assumptions are also imposed on the system settings [7, 12, 13].

**Assumption 2.** *As $N, K \to \infty$, the matrices $\mathbf{D}$ and $\tilde{\mathbf{D}}$ have uniformly bounded spectral norm, i.e.,*

$$\sup ||\mathbf{D}|| < \infty \quad \sup ||\tilde{\mathbf{D}}|| < \infty. \quad (13)$$

*Moreover,*

$$\inf \frac{1}{K} \mathrm{Tr}\mathbf{D} > 0 \quad \inf \frac{1}{K} \mathrm{Tr}\tilde{\mathbf{D}} > 0. \quad (14)$$

*Also, we assume that the family of deterministic $N \times K$ matrices $\mathbf{A}$ has bounded spectral norm:*

$$\sup ||\mathbf{A}|| < \infty. \quad (15)$$

Let us now introduce the fundamental equations that are needed to express a deterministic equivalent of $\gamma_k$. We start with the following set of equations:

$$\delta = \frac{1}{K}\mathrm{Tr}\mathbf{D} \left( \lambda \left( \mathbf{I}_N + \tilde{\delta}\mathbf{D} \right) + \mathbf{A} \left( \mathbf{I}_K + \delta\tilde{\mathbf{D}} \right)^{-1} \mathbf{A}^H \right)^{-1}$$

$$\tilde{\delta} = \frac{1}{K}\mathrm{Tr}\tilde{\mathbf{D}} \left( \lambda \left( \mathbf{I}_K + \delta\tilde{\mathbf{D}} \right) + \mathbf{A}^H \left( \mathbf{I}_N + \tilde{\delta}\mathbf{D} \right)^{-1} \mathbf{A} \right)^{-1}$$

which admits a unique positive solution in the class of Stieltjes transforms of non-negative measures with support $\mathbb{R}_+$ [12, 13]. The matrices

$$\mathbf{T} = \left( \lambda \left( \mathbf{I}_N + \tilde{\delta}\mathbf{D} \right) + \mathbf{A} \left( \mathbf{I} + \delta\tilde{\mathbf{D}} \right)^{-1} \mathbf{A}^H \right)^{-1}$$

$$\tilde{\mathbf{T}} = \left( \lambda \left( \mathbf{I}_K + \delta\tilde{\mathbf{D}} \right) + \mathbf{A}^H \left( \mathbf{I}_N + \tilde{\delta}\mathbf{D} \right)^{-1} \mathbf{A} \right)^{-1}$$

$$\overline{s}_k = \bar{u}_k - \left[ \frac{1}{\lambda \tilde{t}_{kk}^2} \frac{\mathbf{a}_k^H \mathbf{T}^2 \mathbf{a}_k}{(1+\delta \tilde{d}_k)^2} + \left( \frac{1-F}{\Delta}\alpha + \frac{\vartheta}{\Delta}V \right) \left( \lambda \tilde{d}_k + \sum_{i=1, i \neq k}^{K} \frac{\tilde{d}_i}{\lambda \tilde{t}_{k,k}^2} \frac{\left| \mathbf{a}_k^H \mathbf{T} \mathbf{a}_i \right|^2}{\left(1+\delta \tilde{d}_i\right)^2 \left(1+\delta \tilde{d}_k\right)^2} \right) + \right.$$
$$\left. + \left( \frac{1-F}{\Delta}V + \lambda^2 \frac{\tilde{\vartheta}\alpha}{\Delta} \right) \frac{1}{\lambda \tilde{t}_{kk}^2} \frac{\mathbf{a}_k^H \mathbf{T}\mathbf{D}\mathbf{T}\mathbf{a}_k}{(1+\delta \tilde{d}_k)^2} \right] \tag{16}$$

$$\overline{\psi} = \frac{1}{K}\mathrm{Tr}\mathbf{T} - \lambda \left[ \frac{1}{K}\mathrm{Tr}\mathbf{T}^2 + 2\frac{1-F}{\Delta}\alpha V + \frac{\vartheta}{\Delta}V^2 + \frac{\lambda^2 \tilde{\vartheta}\alpha^2}{\Delta} \right]. \tag{17}$$

---

are approximations of the resolvent $\mathbf{Q} = (\mathbf{\Sigma}\mathbf{\Sigma}^H - z\mathbf{I}_N)^{-1}$ and the co-resolvent $\tilde{\mathbf{Q}} = (\mathbf{\Sigma}^H\mathbf{\Sigma} - z\mathbf{I}_K)^{-1}$. Define

$$\vartheta = \frac{1}{K}\mathrm{Tr}\mathbf{D}\mathbf{T}\mathbf{D}\mathbf{T} \tag{18}$$

$$\tilde{\vartheta} = \frac{1}{K}\mathrm{Tr}\tilde{\mathbf{D}}\tilde{\mathbf{T}}\tilde{\mathbf{D}}\tilde{\mathbf{T}} \tag{19}$$

$$F = \frac{1}{K}\mathrm{Tr}\mathbf{T}\mathbf{D}\mathbf{T}\mathbf{A}\left(\mathbf{I}_K + \delta\tilde{\mathbf{D}}\right)^{-2}\tilde{\mathbf{D}}\mathbf{A}^H \tag{20}$$

$$\Delta = (1-F)^2 - \lambda^2 \vartheta\tilde{\vartheta} \tag{21}$$

$$\alpha = \frac{1}{K}\mathrm{Tr}\mathbf{D}\mathbf{T}^2 \tag{22}$$

$$V = \frac{1}{K}\mathrm{Tr}\mathbf{T}^2\mathbf{A}\left(\mathbf{I}_K + \delta\tilde{\mathbf{D}}\right)^{-2}\tilde{\mathbf{D}}\mathbf{A}^H. \tag{23}$$

Also, denote by $\mathbf{a}_k$ the $k$-th column of matrix $\mathbf{A}$ and call $\tilde{t}_{kk}$, $d_k$ and $\tilde{d}_k$ the $k$-diagonal element of $\tilde{\mathbf{T}}$, $\mathbf{D}$ and $\tilde{\mathbf{D}}$, respectively. Then, the following theorem summarizes the main result of this work.

**Theorem 1.** *Let Assumptions 1 – 2 hold true. Then, we have that* $\max_{1 \leq k \leq K} |\gamma_k - \overline{\gamma}_k| \to 0$ *almost surely with*

$$\overline{\gamma}_k = \frac{\bar{u}_k^2}{\overline{s}_k + \overline{\psi}(1+\bar{u}_k)^2 \frac{\sigma^2}{P_T}} \tag{24}$$

*with $\bar{u}_k$ given by*

$$\bar{u}_k = \delta \tilde{d}_k + \frac{1}{\lambda \tilde{t}_{kk}} \frac{\mathbf{a}_k^H \mathbf{T} \mathbf{a}_k}{(1+\delta \tilde{d}_k)} \tag{25}$$

*whereas $\overline{s}_k$ and $\overline{\psi}$ are given on the top of this page.*

*Proof.* The proof is very much involved and relies on results in random matrix theory [12] as well as some recent ones on the deterministic equivalent of bilinear forms [13, Theorem 1]. Due to the space limitations, it is omitted. A complete proof is provided in [20]. □

**Corollary 1.** *Let Assumptions 1 – 2 hold true. If $\mathbf{\Theta} = \mathbf{I}_N$, then $\overline{s}_k$ and $\overline{\psi}_k$ reduce to*

$$\overline{s}_k = \bar{u}_k - \left[ \frac{1-F}{\Delta} \frac{1}{\lambda \tilde{t}_{kk}^2} \frac{\mathbf{a}_k^H \mathbf{T}^2 \mathbf{a}_k}{\left(1+\delta \tilde{d}_k\right)^2} + \frac{\vartheta}{\Delta} \left( \lambda \tilde{d}_k + \right. \right.$$
$$\left. \left. + \sum_{i=1, i \neq k}^{K} \frac{\tilde{d}_i}{\lambda \tilde{t}_{kk}^2} \frac{\left| \mathbf{a}_k^H \mathbf{T} \mathbf{a}_i \right|^2}{\left(1+\delta \tilde{d}_i\right)^2 \left(1+\delta \tilde{d}_k\right)^2} \right) \right] \tag{26}$$

*and*

$$\overline{\psi} = \frac{1}{K}\mathrm{Tr}\mathbf{T} - \frac{\lambda}{\Delta} \frac{1}{K}\mathrm{Tr}\mathbf{T}^2. \tag{27}$$

*Proof.* The result follows easily observing that if $\mathbf{\Theta} = \mathbf{I}_N$ then $F = V = \frac{1}{K}\mathrm{Tr}\mathbf{T}^2\mathbf{A}(\mathbf{I}_K + \delta\tilde{\mathbf{D}})^{-2}\tilde{\mathbf{D}}\mathbf{A}^H$ and $\vartheta = \alpha = \frac{1}{K}\mathrm{Tr}\mathbf{T}^2$. □

Let us now consider the case in which only the link of the user of interest, i.e., UE $k$, is characterized by a Rician fading channel model whereas a Rayleigh fading channel model is considered for all the other UEs with indexes $i \neq k$. Assume also that $||\tilde{\mathbf{z}}_k||^2 = N$ (such as for example when a uniform linear array is used). Then, the following result is obtained.

**Corollary 2.** *Let Assumptions 1 – 2 hold true. If $\mathbf{\Theta} = \mathbf{I}_N$, $||\tilde{\mathbf{z}}_k||^2 = N$ and no LOS component is present $\forall i \neq k$, then $\overline{\gamma}_k$ reduces to*

$$\overline{\gamma}_k = \frac{1}{\left(1 - \lambda\frac{K}{N} \frac{\delta}{1 - \frac{K}{N}\sum_{i=1}^{K}\frac{(\delta\beta_i)^2}{(1+\delta\beta_i)^2}}\right)} \frac{(\delta\beta_k)^2}{\delta\beta_k + \delta\left(1+\delta\beta_k\right)^2 \frac{\sigma^2}{P_T}}$$

*with $\delta$ being the unique positive solution of the following equation:*

$$\delta = \left( \frac{K}{N}\lambda + \frac{1}{N}\sum_{i=1}^{K}\frac{\beta_i}{(1+\delta\beta_i)} \right)^{-1}. \tag{28}$$

The above expression coincides with that in [7] obtained for the case in which all UEs experience Rayleigh fading. This indicates that the asymptotic expression of the SINR for a given user interest is independent from the underlying channel model if all the others are affected by Rayleigh fading and uniform linear array is used.

We are ultimately interested in the individual rates $\{r_k\}$ and the ergodic sum rate $r_E$. Since the logarithm is a continuous function, by applying the continuous mapping theorem, from the almost sure convergence results of Theorem 1 it follows that $r_k - \overline{r}_k \to 0$ almost surely with [7]

$$\overline{r}_k = \log_2\left(1 + \overline{\gamma}_k\right). \tag{29}$$

An approximation of $r_E$ is obtained as follows [7]

$$\overline{r}_E = \sum_{k=1}^{K} \log_2\left(1 + \overline{\gamma}_k\right) \tag{30}$$

such that $\frac{1}{K}(r_E - \overline{r}_E) \to 0$ holds true almost surely.

## 4. NUMERICAL RESULTS

Monte-Carlo simulations are now used to validate the above asymptotic analysis for a network with finite size[1]. We consider a cell of radius $R = 250$ m. We use the standard correlation model $[\mathbf{\Theta}]_{i,j} = \nu^{|i-j|}$ and assume that the large scale coefficients $\beta_k$ are obtained as $\beta_k = 2L_{\bar{x}}(1 + x_k^\kappa/\bar{x}^\kappa)^{-1}$. The parameter $\kappa > 2$ is the path loss exponent, $x_k$ denotes the distance of UE $k$ from the BS, $\bar{x} > 0$ is some cut-off parameter and $L_{\bar{x}}$ is a constant that regulates the attenuation at distance $\bar{x}$. We assume $\kappa = 3.5$, $L_{\bar{x}} = -86.5$ dB and $\bar{x} = 25$ m. The results are obtained for 1000 different channel realizations assuming that the UEs are randomly distributed in the cell. We assume that a uniform linear array is adopted at the BS such that $n$th entry of $\tilde{\mathbf{z}}_k$ is $[\tilde{\mathbf{z}}_k]_n = \exp(-\mathrm{i}(n-1)\pi \sin \theta_k)$ where $\theta_k$ is the arrival angle of UE $k$ taking values in the interval $[-\pi, \pi)$. The transmit power $P_T$ is fixed to 10 Watt and the regularization parameter $\lambda$ is computed as $\lambda = \sigma^2 \mathbb{E}\left[\beta_k^{-1}\right]/P_T$, where the expectation is taken over the distribution of the users' positions.

---
[1]To enable simple testing of other parameter values, the code is also available for download at the following address https://github.com/lucasanguinetti/downlink-Rician-large-scale-MIMO-systems.

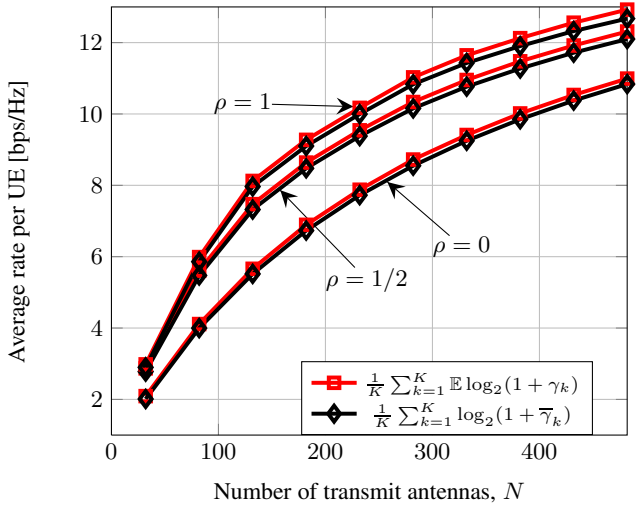

**Fig. 1**. Average rate per UE vs $N$ when $K = 16$ and the Rician factor $\rho$ is $0, 1/2$ and $1$

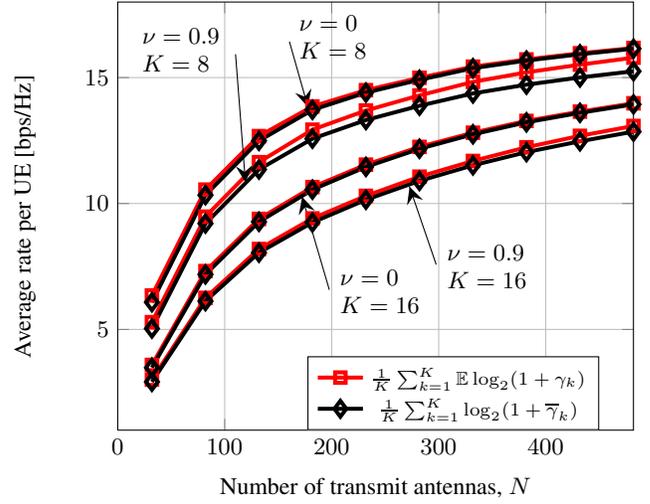

**Fig. 2**. Average rate per UE vs $N$ when $K = 8, 16$ and $\rho = 1$ for different values of the correlation factor $\nu$ among BS antennas.

Fig. 1 illustrates the average per UE rate when $N$ grows large and $K$ is kept fixed to 16. The correlation factor is set to $\nu = 0.9$ and the Rician factor is $\rho = 0, 1/2$ and $1$. As seen, the theoretical asymptotic results match very well with Monte Carlo simulations over the entire of $N$, also when $N$ takes relatively small values. As expected, increasing the Rician factor $\rho$ improves the system performance. The impact of the correlation matrix is analyzed in Fig. 2 for $\rho = 1$ and $K = 8, 16$. The UEs are distributed uniformly on a circle of radius $2/3R$ around the BS. As expected, reducing the correlation factor $\nu$ improves the system performance as it increases the spatial multiplexing capabilities of the channel model.

## 5. CONCLUSION

In this work, we analyzed the ergodic sum rate in the downlink of a single-cell large-scale MIMO system operating over a Rician fading channel. A regularized zero-forcing precoding scheme under the assumption of perfect channel state information is used. Recent results from large-scale random matrix theory allowed us to give concise approximations of the SINRs. Such approximations turned out to depend only on the long-term channel statistics, the Rician factor and the deterministic component. Numerical results indicated that these approximations are very accurate. Applied to practical networks, such results may lead to important insights on how the different parameters affect the performance and allow to simulate the network behavior without the need of extensive Monte Carlo simulations. More details and insights on these aspects will be given in the extended version [20].